\documentclass[a4,12pt]{article}
\input{Qcircuit}
\usepackage{dcolumn}
\usepackage{longtable}

\usepackage{graphicx}
 \usepackage{setspace}
\title{Splitting of quantum information using $N$-qubit linear cluster states}
\author{Sreraman Muralidharan$^{a}$, Sakshi Jain$^{b}$, and \\Prasanta K. Panigrahi$^{c}$\\
$^{a}$ Department of Information and Communication Technology, \\Royal Institute of Technology (KTH), SE-164 40 Kista, Sweden,\\
$^{b}$ Indian Institute of Technology, Mumbai - 400076, India,\\
$^{c}$ Indian Institute of Science Education and Research (IISER)-Kolkata, Mohanpur,\\ BCKV main campus, Nadia-741252, West Bengal, India. \\\\
Address : Prasanta K. Panigrahi\\\\
Indian Institute of Science Education and Research (IISER)-Kolkata, Mohanpur,\\ BCKV main campus, Nadia-741252, West Bengal\\
Phone number : +91-9748918201, Fax : +91-3323348092}
\date{07/02/2010}
\begin{document}
\maketitle
\newpage
\abstract
We provide a number of schemes for the splitting up of quantum information among $k$ parties using a $N$-qubit linear cluster state as a quantum channel, such that the original information can be reconstructed only if all the parties cooperate. Explicit circuits are provided for these schemes, which are based on the concept of measurement based locking and unlocking of quantum information. These are  experimentally feasible as they require measurements to be performed only on product basis. 

\section{Introduction}

Secret sharing between multiple parties, where one or more members can receive a desired message, with the
concurrence of the sender and other members, is a subject of significant current interest. As is evident, this problem is of considerable
importance in the area of intelligence sharing, banking and many other sectors of public interest. The classical methods of secret sharing are
prone to eavesdropping and other forms of tampering, where the involved parties, may not be aware of the presence of Eavesdroppers. The advent of
quantum information \cite{Nielson} and communication has brought in a completely new perspective to this classical problem, wherein not only the channels of
communication can be made secure, but also the presence of eavesdroppers can be detected \cite{Gott}. The fundamental aspects of quantum mechanics, which makes
this possible are, 1) the process of measurement necessarily affects the
state being measured and, 2) the quantum correlations in the communication
channels, arising from  entanglement, an  intrinsic quantum property,
offers unique advantage of detecting tampering and other forms of external
influence. 
The technique of splitting and sharing of quantum information among two or more parties, such that none of them can retrieve the information fully by operating on their own qubits, is usually referred to as Quantum information splitting (QIS). QIS of  $|\psi_1\rangle = \alpha|0\rangle + \beta|1\rangle$, $(\alpha, \beta \in C, |\alpha|^2 + |\beta|^2 = 1)$ has been proposed using $GHZ$ \cite{Hillery, Sam} and asymmetric W states \cite{Zheng}.  Later, QIS of $|\psi_1\rangle$ was experimentally demonstrated using single photon sources \cite{eqis}.

In the recent literaure, a number of quantum networks, with
only a few constituents have been analysed, which has demonstrated, the
in-principle feasibilty of quantum secret sharing and its advantage over,
the classical protocols. In this context, a special class of entangled
channels, the Cluster states \cite{Robert, one, cluster1,cluster2,cluster3}, owing their origin to relatively better understood, Ising spin systems, have
attracted considerable attention recently, because of their ability to
carry out several quantum tasks, in a physically transparent manner \cite{Sreramanc}.

Recently, the search for genuinely entangled channels, which can be used for the deterministic QIS of an arbitrary two qubit state $|\psi_{2}\rangle_{12}=\alpha|00\rangle + \mu|10\rangle + \gamma|01\rangle + \beta|11\rangle$, where $|\alpha|^2+|\mu|^2+|\gamma|^2+|\beta|^2$=1 and $\alpha,$ $\mu,$  $\gamma,$ $\beta \in C$ has attracted much attention. It is worth mentioning that the $GHZ$ and the asymmetric W states cannot be used for the QIS of arbitrary two qubit state $|\psi_{2}\rangle_{12}$, because they do not possess the required entangled structure to carry out the task \cite{Sreramans}. However, a few  specifically entangled multiqubit states \cite{Sreramanb, Sreramanc, Sreraman6, Wang, Sreramanm} have been found useful for splitting $|\psi_2\rangle_{12}$ only among three parties. Further, these protocols required the parties to perform entangled measurements which are extremely difficult to realize in laboratory conditions. It is worth mentioning that one would require atleast five qubits for splitting of an arbitrary two qubit state \cite{Sreramanc}.

This motivates us to devise protocols for the splitting up of $|\psi_2\rangle_{12}$ among $k$ parties, using a $N$ qubit linear cluster state  by utilizing only product basis measurements. In general, an $N$-qubit linear cluster state can be represented as \cite{one}
\begin{equation}
|C_N\rangle = \frac{1}{2^{N/2}} \otimes_{a=1}^{N} (|0\rangle_a \sigma_z ^{a+1} + |1\rangle_a).
\end{equation}
 
The paper has been organized as follows. We first describe explicit circuits for the generation of $N$-qubit linear cluster states. In the next section, we study  the splitting of arbitrary two qubit quantum information $|\psi_2\rangle_{12}$ among $k$ different parties. Explicit circuits for the same have been constructed, wherein the measurements have been performed on the product basis. In the last section, we explain the protocol further by giving illustrations of QIS using five and six qubit cluster states.

In general, any $N$-qubit linear cluster state can be generated from $|000....0\rangle_{123..N}$ by implementing the circuit diagram shown in Fig. 1.
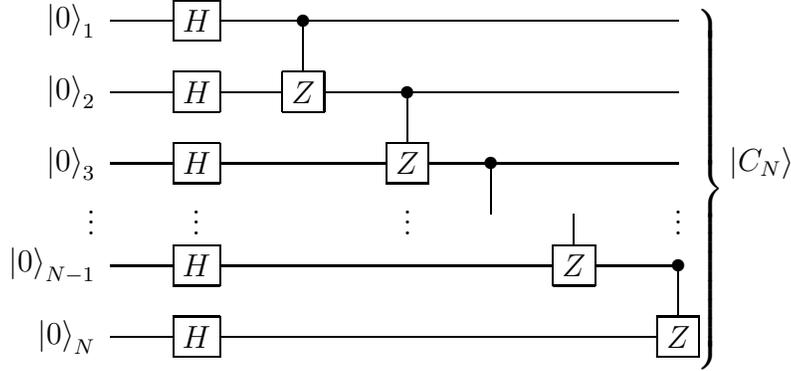
\begin{figure}[h]
	\caption{Circuit diagram for the generation of $|C_N\rangle$}
	\label{fig:CircuitDiagram}
	\leavevmode
\centering	
\Qcircuit @C=2em @R=1em {
\lstick{\ket{0}_1}& \gate{H} &\ctrl{1} &\qw &\qw &\qw &\qw\\
\lstick{\ket{0}_2}& \gate{H} &\gate{Z} \qwx &\ctrl{1} &\qw &\qw &\qw\\
\lstick{\ket{0}_3}& \gate{H} &\qw &\gate{Z} \qwx &\ctrl{1} &\qw &\qw & |C_{N}\rangle\\
\lstick{\vdots}& \vdots & &\vdots & & &\vdots\\
\lstick{\ket{0}_{N-1}}& \gate{H} &\qw &\qw &\qw &\gate{Z} \qwx &\ctrl{1}\\
\lstick{\ket{0}_N}& \gate{H} &\qw &\qw &\qw &\qw &\gate{Z} \qwx
\ \gategroup{1}{7}{6}{7}{.7em}{\}} }  \\
\end{figure}

\section{QIS of $|\psi_2\rangle_{12}$ among $k$ parties}
The protocol for the splitting of an arbitrary two qubit secret $|\psi_2\rangle_{12}$ among $k$ different parties using $|C_N\rangle$ can be divided into two major steps: "Locking" and "unlocking" of quantum secret. We label the participants Alice, $Bob_{1}$, $Bob_{2}$,... $Bob_{k-1}$ and Charlie, where Charlie is designated to get the final state.  Before distributing the qubits among the parties, the qubits of $|C_N\rangle$ are swapped in the following manner,

\begin{eqnarray}
|C_{N}\rangle \stackrel{Swap(N-2,N), . . . , Swap(3,5),Swap(1,3)} {\rightarrow} (|C^{'}_N\rangle), \text{if N is odd} \\
|C_{N}\rangle \stackrel{Swap{(N/2,N)}, Swap{1,(N/2 + 1)}, . . ., Swap(2,4)} {\rightarrow} (|C^{'}_N\rangle), \text{if N is even}
\end{eqnarray}
where $Swap{(i,j)}$ represents the swapping of "i"th and "j"th qubits respectively. We now distribute the qubits such that $|c_{1}\rangle$ and $|c_{2}\rangle$ belong to Alice, $|c_{3}\rangle$ and $|c_{4}\rangle$ belong to $Bob_{1}$, qubit $|c_{5}\rangle$ to $Bob_2$, .. and the qubits $|c_{(N-1)}\rangle$ and $|c_{N}\rangle$ to Charlie, where the "$i$"th qubit ($i \geq N$) of $|C_{N}\rangle$ is denoted by $|c_{i}\rangle$. 
The QIS scheme for $N=5$ and $6$ will be explicated below. For $N \geq 6$, we let Alice, $Bob_1$, Charlie  possess two qubits each and each of the remaining $(N-5)$ participants possess one qubit. 
\subsection{Locking the quantum secret}
In order to lock $|\psi\rangle_{12}$ among the other participants, she initially swap the qubits of $|C_N\rangle$ as per the rule discussed above and swaps the qubit $|\psi_2\rangle_{2}$ and $|C_N\rangle_{2}$, as is explicitly shown in Fig.2. This is followed by a $CNOT$ gate between $|\psi_1\rangle$ and $|\psi_2\rangle$, a Hadamard on $|\psi_2\rangle$ in order to "break" the entangled measurements into product measurements.  She measures each of her four qubits individually in the basis ${(|0\rangle, |1\rangle)}$ and conveys the outcome of the measurement to Charlie via four classical bits. The information is thus locked amongst the parties Bob$_{1}$,  Bob$_{2}$ ... Bob$_{(N-5)}$,  $(N>5)$ and Charlie such that none of them can obtain the quantum secret by operating on their own qubits.  The circuits explicitly constructed for this protocol are shown in Fig. 2 or Fig. 3 for a even or odd $N$ respectively.

\subsection{Unlocking the quantum secret} 
For unlocking $|\psi_2\rangle_{12}$, the parties should act as follows. Initially, $Bob_1$ performs a $CNOT_{3,4}$ operation on the two qubits $|c_{3}\rangle$ and $|c_{4}\rangle$, projects the two qubits on the computational basis given by, {($|00\rangle$, $|01\rangle$, $|10\rangle$, $|11\rangle$)} and conveys the outcome of the measurement to Charlie via two cbits.  The other participants, $Bob_{i}, i \in {2,...(N-5)}$ perform a Hadamard measurement $\frac{1}{\sqrt{2}}(|0\rangle\pm |1\rangle)$ and conveys the outcome to Charlie via cbits. Once Charlie obtains all the ($N-4$) measurement results (including Alice's measurement outcome), he can perform a suitable set of operations on his two qubits and deterministically obtain $|\psi_2\rangle_{12}$. Thus, Alice's quantum secret $|\psi_2\rangle_{12}$ which was initially split among ($N-5$) intermediate parties was sent to Charlie by performing only product basis measurements. This completes the proposed QIS scheme. The quantum circuits in Fig. 2 and Fig. 3, depending on whether $N$ is even or odd, show these steps clearly. 

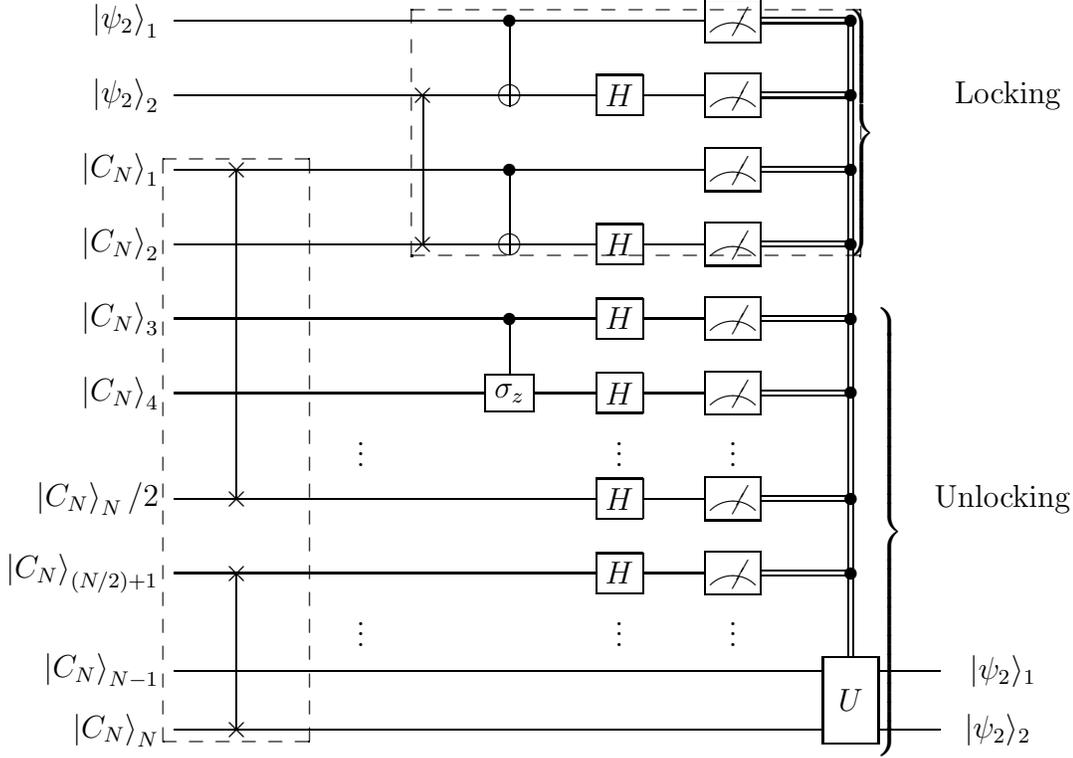
\begin{figure}[h]
	\caption{Circuit diagram for the locking and unlocking in QIS using $|C_{N}\rangle$ ($N$ is even)}
	\label{fig:CircuitDiagram}
	\leavevmode
\centering	
\Qcircuit @C=2em @R=1em {
\lstick{\ket{\psi_2}_1} &\qw &\qw &\qw &\qw &\ctrl{1} &\qw &\meter &\control \cw &\\
\lstick{\ket{\psi_2}_2} &\qw &\qw &\qw &\qswap &\targ &\gate{H} &\meter&\control \cw\cwx & &\text{   Locking}  \\
\lstick{\ket{C_{N}}_1} &\qswap &\qw &\qw &\qw \qwx &\ctrl{1} &\qw &\meter&\control \cw\cwx &  \\
\lstick{\ket{C_{N}}_2} &\qw \qwx &\qw &\qw &\qswap \qwx &\targ &\gate{H} &\meter&\control \cw\cwx &\\
\lstick{\ket{C_{N}}_3} &\qw \qwx &\qw & \qw &\qw &\ctrl{1} &\gate{H} &\meter &\control \cw\cwx &  \\
\lstick{\ket{C_{N}}_4} &\qw \qwx &\qw & \qw &\qw &\gate{\sigma_z} \qwx &\gate{H} &\meter &\control \cw\cwx &  \\
\lstick{}                 & \qwx & & \vdots & & &\vdots  &\vdots &\cwx & & & &\\
\lstick{\ket{C_{N}}_N/2} &\qswap \qwx &\qw & \qw &\qw &\qw &\gate{H} &\meter &\control \cw\cwx & &\text{Unlocking} \\
\lstick{\ket{C_{N}}_{(N/2)+1}} &\qswap  &\qw & \qw &\qw &\qw &\gate{H} &\meter &\control \cw\cwx & &  \\
\lstick{}                 & \qwx & &\vdots  & & &\vdots &\vdots & \cwx& & &\\
\lstick{\ket{C_{N}}_{N-1}} &\qw \qwx & \qw &\qw &\qw &\qw &\qw &\qw &\multigate{1}{U} \cwx &\qw &|\psi_2\rangle_{1}  \\
\lstick{\ket{C_{N}}_{N}} &\qswap \qwx &\qw &\qw &\qw &\qw &\qw &\qw & \ghost{U}  & \qw & |\psi_2\rangle_{2} 
\gategroup{3}{1}{12}{3}{.7em}{--} \gategroup{1}{5}{4}{9}{.7em}{--}
\ \gategroup{1}{9}{4}{9}{.7em}{\}} \gategroup{5}{9}{12}{9}{.7em}{\}} }  \\
\end{figure}

 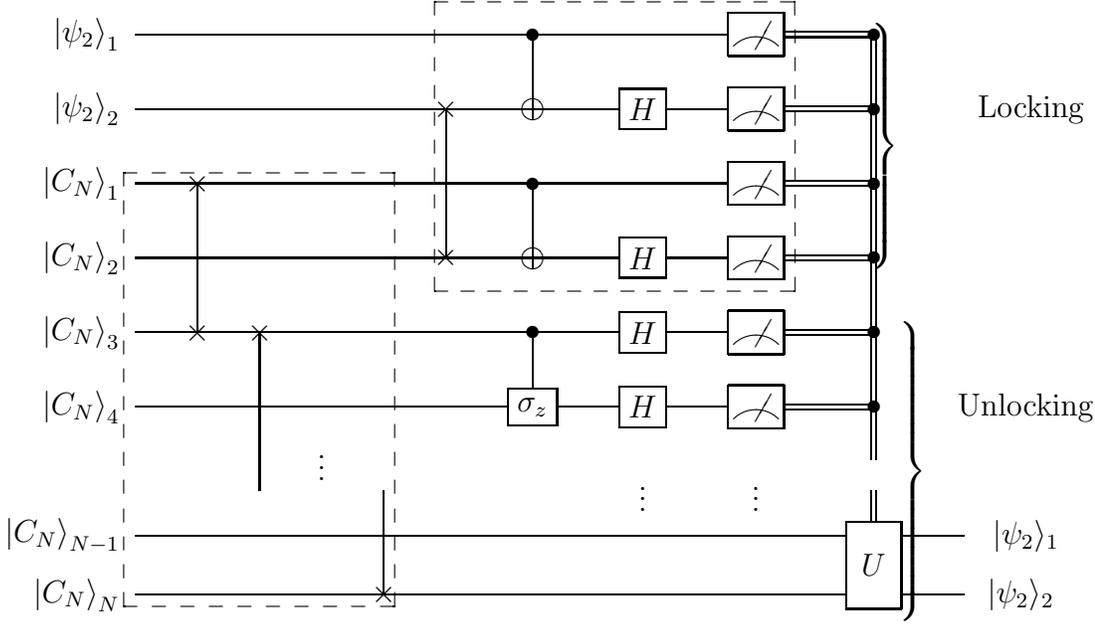
\begin{figure}[h]
	\caption{Circuit diagram for the locking and unlocking in QIS using $|C_{N}\rangle$ ($N$ is odd)}
	\label{fig:CircuitDiagram}
	\leavevmode
\centering	
\Qcircuit @C=2em @R=1em {
\lstick{\ket{\psi_2}_1} &\qw &\qw &\qw &\qw &\qw &\ctrl{1} &\qw &\meter &\control \cw &\\
\lstick{\ket{\psi_2}_2} &\qw &\qw &\qw &\qw &\qswap &\targ &\gate{H} &\meter&\control \cw\cwx &&\text{   Locking}  \\
\lstick{\ket{C_{N}}_1} &\qswap &\qw &\qw &\qw &\qw \qwx &\ctrl{1} &\qw &\meter&\control \cw\cwx &  \\
\lstick{\ket{C_{N}}_2} &\qw \qwx &\qw &\qw &\qw &\qswap \qwx&\targ &\gate{H} &\meter&\control \cw\cwx & \\
\lstick{\ket{C_{N}}_3} &\qswap \qwx &\qswap & \qw &\qw &\qw &\ctrl{1} &\gate{H} &\meter &\control \cw\cwx &  \\
\lstick{\ket{C_{N}}_4} &\qw &\qw \qwx & \qw &\qw &\qw &\gate{\sigma_z} \qwx &\gate{H} &\meter &\control \cw\cwx && \text{Unlocking}  \\ 
\lstick{}                 & &\qwx & \vdots & & & & & &\cwx & & & &\\
\lstick{}                 & &\qwx &  & & & &\vdots &\vdots & & & &\\
\lstick{\ket{C_{N}}_{N-1}} &\qw & \qw &\qw &\qw \qwx&\qw &\qw &\qw &\qw &\multigate{1}{U} \cwx &\qw & |\psi_2\rangle_{1}\\
\lstick{\ket{C_{N}}_{N}} &\qw &\qw &\qw &\qswap \qwx &\qw &\qw &\qw &\qw & \ghost{U}  & \qw & |\psi_2\rangle_{2} 
\gategroup{3}{1}{10}{5}{.7em}{--} \gategroup{1}{6}{4}{9}{.7em}{--}
\ \gategroup{1}{10}{4}{10}{.7em}{\}} \gategroup{5}{10}{10}{10}{.7em}{\}} }  \\
\end{figure}

\section{Illustrations}
We shall now illustrate the above proposed protocol explicitly for $N = 5$ and $N = 6$ respectively. We shall also provide relations between the classical bits received by Charlie by the different parties and the local operations to be performed by him in order to deterministically obtain $|\psi_2\rangle_{12}$. 

\subsection{QIS of $|\psi_2\rangle_{12}$ using five qubit cluster state $|C_5\rangle$}
The five qubit  cluster state  
\begin{equation}
|C_5\rangle = \frac{1}{2}(|00101\rangle - |00010\rangle - |11001\rangle + |11110\rangle), 
\end{equation}
can be generated using the circuit shown in Fig. 1.
After performing the required SWAP operations between qubits 1 and 3 and the qubits 3 and 5, the resultant state is given by, 
\begin{equation}
|C^{'}_5\rangle = \frac{1}{2}(|00010\rangle + |01101\rangle - |10100\rangle - |11011\rangle).
\end{equation}
$|C'_5\rangle$ forms an important resource for QIS among three parties. The qubits are distributed such that Alice possesses the qubits 1 and 2 of $|C'_5\rangle$ along with $|\psi_{2}\rangle_{12}$, which is to be split among the two parties, $Bob_{1}$ and Charlie. We let $Bob_{1}$ possess qubit 3 and Charlie possess qubits 4 and 5. In the next step, Alice performs a measurement on each of her four qubits individually in the basis ($|0\rangle$, $|1\rangle$), thereby locking the quantum secret in the Bob-Charlie system. She then conveys the outcome of her measurement to Charlie via four classical bits. It is worth mentioning that, at this stage Charlie cannot decipher $|\psi_2\rangle_{12}$, with Alice's measurement outcome alone. In order to unlock $|\psi_{2}\rangle_{12}$, $Bob_1$ performs a Hadamard measurement on his qubit (since no entangling operation is performed for $N \leq 5$) and sends the result to Charlie via one classical bit. Having obtained the outcomes of both Alice and $Bob_{1}$, Charlie can now deterministically reconstruct $|\psi_{2}\rangle_{12}$ by applying suitable unitary operations on his qubits.

We denote the $4$ classical bits sent by Alice to Charlie as, "$a_1a_2a_3a_4$" and the single classical bit sent by $Bob_1$ as ''$b_1$". The unitary local operation $U$ to be performed by Charlie in order to obtain $|\psi_{2}\rangle_{12}$ is then given by,
\begin{flalign}
U = (\overline{a_4}.\overline{a_2}(\sigma_x \otimes I) + \overline{a_4}.a_2(I \otimes \sigma_x) + a_4.\overline{a_2}(I \otimes I) + a_4.a_2(\sigma_x \otimes \sigma_x)).\\ \nonumber
CNOT_{2,1}.Swap_{1,2}.(\overline{(a_1 \oplus a_3)}.\overline{(a_3 \oplus b_1)}(I \otimes \sigma_z) + \overline{(a_1 \oplus a_3)}.(a_3 \oplus b_1)(\sigma_z \otimes I)\\ \nonumber
 + (a_1 \oplus a_3).\overline{(a_3 \oplus b_1)}(\sigma_z \otimes \sigma_z) + (a_1 \oplus a_3).(a_3 \oplus b_1)(I \otimes I) ).
\end{flalign}
Here, $\oplus$ and $\overline{a_i}$ denote the classical XOR and NOT respectively.  

For instance, let us suppose that the cbits sent by Alice, $a_1a_2a_3a_4$ be $1110$, and that by Bob be "$1$". The unitary operation $U$ that Charlie should apply on his two qubits is then given by, $U = (\sigma_x \otimes I).CNOT_{2,1}.Swap_{1,2}.(I \otimes \sigma_z)$. The local operations $U$ for other classical messages can be obtained in a similar manner.

An explicit circuit showing the two stages, locking and unlocking of the $|\psi_{2}\rangle_{12}$ is given in Fig. 4.
This protocol assumes significance, since five is the threshold number of qubits that is required for the
QIS of an arbitrary two qubit state $|\psi_2\rangle_{12}$ in the case where both the parties involved need not meet.
Further, this protocol is easier for experimental implementation than the previous protocol as it involves only product measurements \cite{Sreramanc}. 

 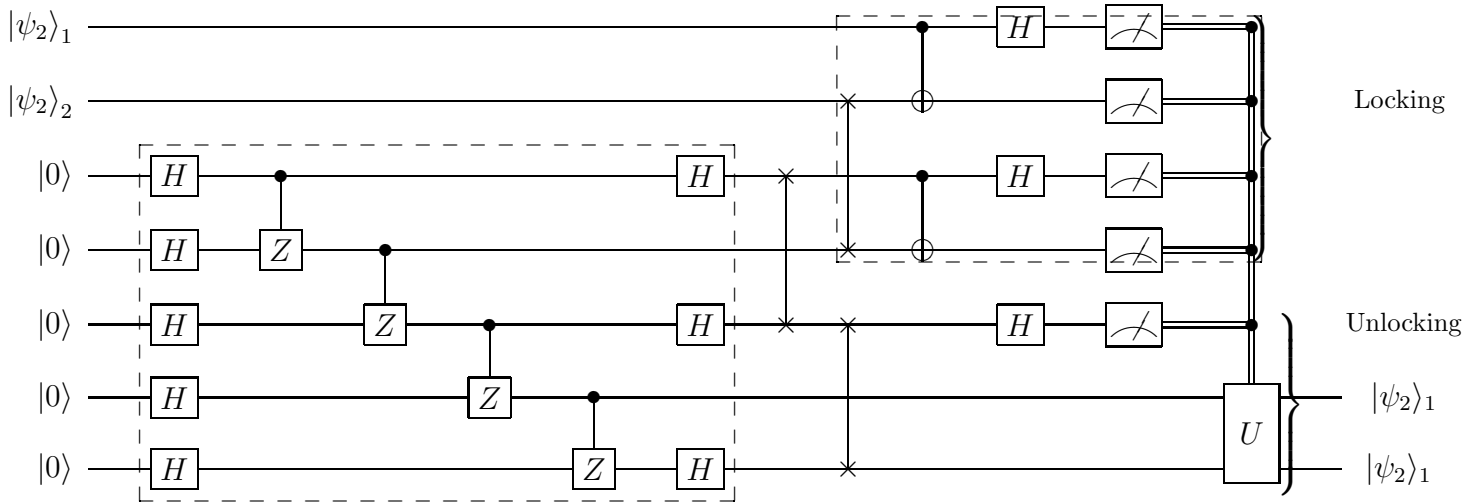
\begin{figure}[h]
	\caption{Circuit diagram for the locking and unlocking in QIS using $|C_5\rangle$}
	\label{fig:CircuitDiagram}
	\leavevmode
\centering	
\Qcircuit @C=2em @R=1em {
\lstick{\ket{{\psi_2}}_1} &\qw &\qw &\qw &\qw &\qw &\qw &\qw & \qw &\ctrl{1} &\gate{H} &\meter &\control \cw \\
\lstick{\ket{{\psi_2}}_2} &\qw &\qw &\qw &\qw &\qw &\qw &\qw &\qswap &\targ \qwx &\qw &\meter&\control \cw\cwx &&\mbox{\fontsize{10}{12}\selectfont 	Locking }  \\
\lstick{\ket{0}}& \gate{H} &\ctrl{1} &\qw &\qw &\qw &\gate{H} &\qswap &\qw \qwx &\ctrl{1} &\gate{H} &\meter&\control \cw\cwx  \\
\lstick{\ket{0}}& \gate{H} &\gate{Z} \qwx &\ctrl{1} &\qw &\qw &\qw &\qw \qwx &\qswap \qwx &\targ \qwx &\qw &\meter&\control \cw\cwx  \\
\lstick{\ket{0}}& \gate{H} &\qw &\gate{Z} \qwx &\ctrl{1} &\qw &\gate{H} &\qswap \qwx &\qswap &\qw & \gate{H}  &\meter &\control \cw\cwx & &\mbox{\fontsize{10}{12}\selectfont 	Unlocking}\\ 
\lstick{\ket{0}}& \gate{H} &\qw &\qw &\gate{Z} \qwx &\ctrl{1} &\qw &\qw &\qw \qwx&\qw &\qw &\qw &\multigate{1}{U} \cwx &\qw  &|\psi_2\rangle_{1}\\
\lstick{\ket{0}}& \gate{H} &\qw &\qw &\qw &\gate{Z} \qwx &\gate{H} &\qw &\qswap \qwx &\qw &\qw  &\qw & \ghost{U}  & \qw &|\psi_2\rangle_{1}
\gategroup{3}{2}{7}{7}{.7em}{--} \gategroup{1}{9}{4}{13}{.7em}{--}
\ \gategroup{1}{13}{4}{13}{.7em}{\}} \gategroup{5}{13}{7}{13}{.7em}{\}} }  \\
\end{figure}

\subsection{QIS of $|\psi_2\rangle_{12}$ using six qubit cluster state}
The  six qubit linear cluster state $|C_6\rangle$  can be generated using the circuit shown in Fig.1. We then perform swap operations ($Swap(1,4), Swap(3,6)$) on $|C_6\rangle$   and the resultant cluster state is given by,

\begin{flalign}
|C^{'}_6\rangle =\frac{1}{2\sqrt{2}}(|010101\rangle - |010010\rangle - |001001\rangle + |001110\rangle \\ \nonumber
              + |100101\rangle - |100010\rangle - |111001\rangle - |111110\rangle).
\end{flalign}

This state can be used to establish the QIS protocol among $(N-3)=3$ parties namely, Alice, $Bob_{1}$, and Charlie. To initialize the protocol,  we let Alice possess the qubits 1 and 2 (along with $|\psi_{2}\rangle$), $Bob_{1}$ posses qubits 3 and 4 and Charlie possess qubits 5 and 6, as stated in the generalized scheme discussed in section 3.1. Next, Alice performs a four particle computational basis measurement, and conveys the outcome to Charlie using four classical bits, thereby locking the quantum secret between $Bob_{1}$ and Charlie. In the next step, in order to unlock $|\psi_{2}\rangle_{12}$, $Bob_{1}$ performs a $CNOT_{3,4}$ operation on his two qubits. He measures the outcome in the computational basis as in the previous case after applying Hadamard states and sends the results to Charlie via two classical bits.  Having known the outcomes of the measurement of Alice and Bob, Charlie can apply suitable unitary operation $U$ and deterministically retrieve $|\psi_{2}\rangle_{12}$. 

If $a_1a_2a_3a_4$ denotes the four cbits sent by Alice and $b_1b_2$ denote the ones sent by Bob, then the local unitary operation to be performed by Charlie, corresponding to the different messages, is given by,
\begin{eqnarray}
U &=&[\overline{a_4}.((\sigma_x \otimes I).(\overline{(a_1 \oplus a_2 \oplus b_1) \oplus (a_3 \oplus b_2)}) + (I \otimes I).((a_1 \oplus a_2 \oplus b_1) \oplus (a_3 \oplus b_2))) \\ \nonumber
&+& a_4.((I \otimes \sigma_x).(\overline{(a_1 \oplus a_2 \oplus b_1) \oplus (a_3 \oplus b_2)}) \\ \nonumber
&+&(\sigma_x \otimes \sigma_x).((a_1 \oplus a_2 \oplus b_1) \oplus (a_3 \oplus b_2)))].CNOT_{2,1}.[\overline{(a_1 \oplus a_3)}.((\sigma_z \otimes I).(\overline{(a_3 \oplus b_2)}) \\ \nonumber
&+&(I \otimes \sigma_z).(a_3 \oplus b_2)) + (a_1 \oplus a_3).((\sigma_z \otimes \sigma_z).(\overline{(a_3 \oplus b_2)}) + (I \otimes I).(a_3 \oplus b_2))].
\end{eqnarray}
For instance, if the four cbits sent by Alice, $a_1a_2a_3a_4$ are $0100$ and that by Bob $b_1b_2$ are $01$, then the unitary operation $U$ is given by, $U = (I \otimes \sigma_x).CNOT_{2,1}.(I \otimes \sigma_z)$.
Appropriate local operations corresponding to other messages can be obtained in a similar way.

An explicit circuit showing locking and unlocking of the $|\psi_{2}\rangle_{12}$ for the case of $|C_6\rangle$ is shown below in Fig. 5.
 \begin{figure}[h]
	\caption{Circuit diagram for the locking and unlocking of two qubit quantum secret using $|C_6\rangle$}
	\label{fig:CircuitDiagram}
	\leavevmode
\centering	
\Qcircuit @C=2em @R=1em {
\lstick{\ket{{\psi_2}}_1} &\qw &\qw &\qw &\qw &\qw &\qw &\qw &\qw &\qw &\ctrl{1} &\gate{H} &\meter & \control \cw &\\
\lstick{\ket{{\psi_2}}_2} &\qw &\qw &\qw &\qw &\qw &\qw  &\qw &\qw &\qswap &\targ \qwx &\qw &\meter &\control  \cw \cwx & &\mbox{\fontsize{10}{12}\selectfont 	Locking }\\
\lstick{\ket{0}}& \gate{H} &\ctrl{1} &\qw &\qw &\qw &\qw &\gate{H} &\qswap &\qw \qwx &\ctrl{1} &\gate{H} &\meter &\control  \cw \cwx &\\
\lstick{\ket{0}}& \gate{H} &\gate{Z} \qwx &\ctrl{1} &\qw &\qw &\qw &\qw &\qw \qwx &\qswap \qwx &\targ \qwx &\qw &\meter &\control  \cw \cwx &\\
\lstick{\ket{0}}& \gate{H} &\qw &\gate{Z} \qwx &\ctrl{1} &\qw &\qw &\qw  &\qswap \qwx &\qw &\ctrl{1} &\gate{H} &\meter &\control  \cw \cwx &\\
\lstick{\ket{0}}& \gate{H} &\qw &\qw &\gate{Z} \qwx &\ctrl{1} &\qw &\gate{H}  &\qswap &\qw &\gate{\sigma_z} \qwx &\gate{H} &\meter &\control  \cw \cwx &&& \mbox{\fontsize{10}{12}\selectfont Unlocking }\\
\lstick{\ket{0}}& \gate{H} &\qw &\qw &\qw &\gate{Z} \qwx &\ctrl{1} &\qw &\qw \qwx &\qw &\qw &\qw & \multigate{1}{U} \cwx &\qw &\qw &|\psi_2\rangle_{1}\\
\lstick{\ket{0}}& \gate{H} &\qw &\qw &\qw &\qw &\gate{Z} \qwx &\gate{H} &\qswap \qwx &\qw &\qw &\qw &\ghost{U} &\qw &\qw &|\psi_2\rangle_{2}
\gategroup{3}{2}{8}{7}{.7em}{--} \gategroup{1}{10}{4}{14}{.7em}{--}
\ \gategroup{1}{14}{4}{14}{.7em}{\}} \gategroup{5}{14}{8}{15}{.7em}{\}} }  \\
\end{figure}
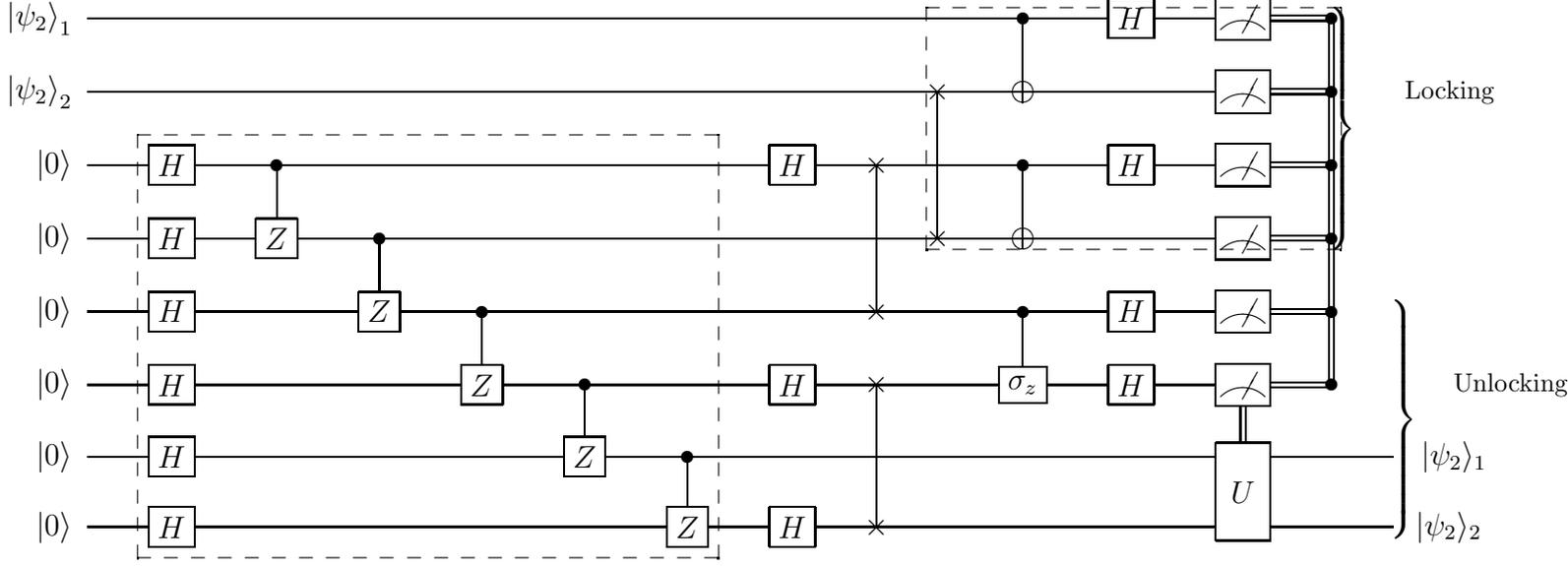

\section{Conclusion}

In this paper, we have explicated the creation and the use of $N$-qubit cluster states for the generalization of quantum information splitting ( QIS )  protocol among $k$ different parties. Explicit circuit diagrams involving only experimentally realizable quantum gates have been described. Unlike the presented scheme, most of the schemes that deal with the QIS of an arbitrary two qubit state in the literature involve splitting of quantum information only among limited number of parties. However, using the protocol proposed in this paper, one can have a QIS scheme that involves any number of parties. Secondly, the schemes developed so far in literature, either involve highly correlated multipartite  measurements, which are extremely difficult to implement in experimental conditions or they use realizable  Bell type measurements but with larger number of entangled photons. However, the illustrated protocol uses $N$-qubit linear cluster states  $|C_{N}\rangle$ which employs only computational basis measurements thereby making it feasible for experimental realization. Further, the initial resource used, i.e., the cluster states are shown to be robust against decoherence \cite{clusterdeco}. We hope that this will lead to experimental realization of QIS of an arbitrary two qubit state among any number of involved parties, which has long been a challenge to the experimentalists.

\end{document}